\documentclass[final,3p,times,onecolumn]{elsarticle}

\usepackage{graphicx}
\usepackage{epsfig}
\usepackage{amsmath,amsfonts, amsthm, mathrsfs, amssymb, bbm}
\usepackage{mathtools}
\usepackage{commath}
\usepackage[sc,osf]{mathpazo}
\usepackage[noend, ruled,vlined,noresetcount]{algorithm2e}
\usepackage{natbib}
\usepackage{times}
\usepackage{subfigure}
\usepackage{lineno}
\usepackage{caption}
\usepackage{amssymb}

\usepackage{multirow}
\allowdisplaybreaks
\journal{arXiv}

\begin{document}

\begin{frontmatter}

\title{The satisficing secretary problem: when closed-form solutions meet simulated annealing}

\author[DMath]{Roberto Brera}
\ead{roberto.brera.24@dartmouth.edu}
\author[DMath,DMed]{Feng Fu\corref{ff}}
\ead{fufeng@gmail.com}

\address[DMath]{Department of Mathematics, Dartmouth College, Hanover, NH 03755, USA}

\address[DMed]{Department of Biomedical Data Science, Geisel School of Medicine at Dartmouth, Lebanon, NH 03756, USA}

\cortext[ff]{Corresponding author at: 27 N. Main Street, 6188 Kemeny Hall, Department of Mathematics, Dartmouth College, Hanover, NH 03755, USA. Tel.: +1 (603) 646 2293}

\begin{abstract}
The secretary problem has been a focus of extensive study with a variety of extensions that offer useful insights into the theory of optimal stopping. The original solution is to set one stopping threshold that gives rise to an immediately rejected sample $r$ out of the candidate pool of size $n$ and to accept the first candidate that is subsequently interviewed and bests the prior $r$ rejected. In reality, it is not uncommon to draw a line between job candidates to distinguish those above the line vs those below the line. Here we consider such satisficing sectary problem that views suboptimal choices (finding any of the top $d$ candidates) also as hiring success. We use a multiple stopping criteria $(r_1, r_2, \cdots, r_d)$ that sequentially lowers the expectation when the prior selection criteria yields no choice. We calculate the probability of securing the top $d$ candidate in closed form solutions which are in excellent agreement with computer simulations. The exhaustive search for optimal stopping has to deal with large parameter space and thus can quickly incur astronomically large computational times. In light of this, we apply the effective computational method of simulated annealing to find optimal $r_i$ values which performs reasonably well as compared  with the exhaustive search despite having significantly less computing time. Our work sheds light on maximizing the likelihood of securing satisficing (suboptimal) outcomes using adaptive search strategy in combination with computation methods. 

\end{abstract}


\end{frontmatter}


\section{Introduction}
\label{intro}

The secretary problem presents a conceptually simple yet mathematically elegant model for studying optimal decision-making strategies that can maximize the probability of finding the best choice (or match) in a range of real-world scenarios, ranging from getting the best job candidate to finding the best partner. Since this now well-known problem first appeared in the February 1960 issue of \textit{Scientific American}, substantial progress concerning the solution and its different variations has been made~\cite{Freeman,ferguson1989solved,samuels1989solved}. Over past decades, systematic efforts on the secretary problem have greatly helped spur the interest from diverse fields with broad implications for informing practices aimed at finding ``the right one(s)'' with various constraints~\cite{gianini1976infinite,corbin1980secretary,stewart1981secretary,abdel1982secretary,rogerson1987probabilities,seale2000optimal,lee2005decision,bearden2006new,mahdian2008secretary,szajowski2016shelf,bradac2019robust,correa2021secretary,chin2021ordinal,bayon2022multi}. Besides, the secretary problem has found wide applications to online trading~\cite{koutsoupias2018online}, online auction~\cite{harrell2015online}, and customer behavior~\cite{zhao2017s}.   

In 1961, Lindley published the first solution to the standard problem~\cite{Lindley}. Since then, the secretary problem has been extended and generalized in numerous ways, including, among others, interview cost~\cite{bartoszynski1978secretary,lorenzen1981optimal}, finite memory length~\cite{rubin1977finite,Smith2}, uncertainty in employment~\cite{Smith1}, observation errors~\cite{skarupski2020secretary}, and repeated secretary problem ~\cite{goldstein2017learning}. The problem has also been studied from various mathematical perspectives, such as via linear programming~\cite{Buchbinder}, submodular functions~\cite{bateni2013submodular}, random walks~\cite{hlynka1988secretary} and graph theoretic approach~\cite{kubicki2005graph}.

More specifically, one important line of extensions allows multiple choice~\cite{mori1984random,preater1994multiple}, or in general $d$ choice \cite{Glasser}.
Chow has worked on a selection model minimizing the rank of the expected selected candidate to $\approx3.87$~\cite{Chow}, and Smith has elaborated on the case of uncertain employment~\cite{Smith1}, where each candidate accepts an offer only with a fixed probability of $p$. Furthermore, Buchbinder, Jain, and Singh have worked on the $J$-choice $K$-best secretary problem~\cite{Buchbinder}, where the decision maker is allowed the selection of $J$ candidates and needs to choose a candidate amongst the top $K$ to achieve success. Other multiple-selection variations have been explored by Albers~\cite{Albers} (where the decision maker must maximize the qualitative sum of a given number of selections) and Glasser~\cite{Glasser}, who computes the probabilities of selecting all the $d$ best candidates when given $d$ possibilities. Smith relaxed different criteria by allowing, at any time, the availability of the last $m\geq1$ interviewed candidates for selection, whilst maintaining all other standard rules~\cite{Smith2}. These variations of the problem, together with several others, have also been studied by Freeman~\cite{Freeman} and Samuels~\cite{Samuels}.

Built on prior studies, the present work concerns a variation of the secretary problem. In the original problem, we are required to choose the best candidate out of a pool of $n$ candidates, where each can be interviewed only once and must be immediately accepted or rejected. Our variation of the problem will have as additional input a tolerance level $d$, and success in the problem is dictated by choosing any of the top $d$ candidates out of the total $n$ candidates.

To inform our approach to the generalized secretary problem, we first look to the standard problem's solution, which is also known as the optimal stopping problem. In order to maximize the likelihood of choosing the best candidate, we implement a strategy where the first $r-1$ candidates are automatically rejected (to obtain a sample of the pool), and then the first candidate which bests the rejected group is accepted. We can compute the $r$ value maximizing the probability of success in selecting the best candidate among the pool.

For a specific value of $r$, this probability is given by:

\begin{align*}
P(r) &= \sum_{i=r}^{n}P( \text{$i$ is the best candidate}) \cdot P(\text{success at selecting $i$ } | \text{ $i$ is the best candidate} ) \\
     &= \sum_{i=r}^{n} {\frac{1}{n}}  \cdot P(\text{best candidate before $i$ lies in the $r-1$ group}) \\
     &= \frac{1}{n}\sum_{i=r}^{n}\frac{r-1}{i-1}\\
     &= \frac{r-1}{n} \sum_{i=r}^n \frac{1}{i-1}.
\end{align*}

As $n \to \infty$, we can approximate the summation as an integral, with $x = \lim_{n \to \infty} \frac{r-1}{n}, t = \lim_{n \to \infty} \frac{i-1}{n},$ and $dt = \lim_{n \to \infty} \frac{1}{n}$. Then, we can take the derivative with respect to $x$ to find the approximate value of $x \approx \frac{r-1}{n}$ for which $P(x)$ is maximized:
\begin{equation*}
    P(x) = x \int_{x}^1 \frac{1}{t} dt = x \ln(x)
\end{equation*}
\begin{equation*}
    \frac{d}{dx} P(x) = \ln(x) + 1 = 0.
\end{equation*}
\vspace{1ex}
Thus, the optimal choosing strategy has probability of success at about  $e^{-1}$ for large $n$  and is achieved when $r \approx \left\lfloor\frac{n}{e}\right\rfloor +1$.

In contrast to this standard case with strict criteria of success (only finding the best candidate), it is not uncommon that the job search is considered a success as long as any of the top $d$ candidates is found. We focus our present study on this ``satsificing'' secretary problem. We propose an adaptive search strategy with multiple stopping criteria $(r_1, r_2, \cdots, r_d)$. For each interval $[r_s, r_{s+1})$, the interviewer (decision-maker) lowers their expectation subsequently aiming to find the $s$th-best candidate relative the prior $r_s$ rejected if prior search throughout the interval $[r_{s-1}, r_{s})$  yields no selection. We are able to derive the probability of success, finding any of the top $d$ admissible candidates, in closed form for any given parameter choices of $r_i$'s values, for $i =  1, \cdots, d$.

In what follows, the paper is organized as follows. Section 2 presents the model details of the satisficing secretary problem. We first present the simple case $d = 2$ which gives us heuristics for  deriving closed-form solutions for the general case $d \ge 2$. We also compare analytical results with computer simulations. Section 3 presents specific results on the optimization of multiple stopping criteria $(r_1, \cdots, r_d)$ using the effective computational method of simulated annealing as compared to the brute force method. We conclude and discuss the work in Section 4.

\section{Model and methods}

\subsection{Satisficing secretary problem with top $d$ admissible candidates}

\vspace{1ex}

When relaxing success criteria to the selection of one amongst a set of $d$ equally admissible candidates, relaxed selection criteria yield a greater probability of success. Thus, we implement a strategy where the first $r_{1}-1$ candidates are automatically rejected, and then, for a set of stopping points $(r_{1},r_{2},...,r_{d})$, between each $r_{i}, r_{i+1}-1$ inclusive we are willing to accept one amongst the best $i$ candidates amongst the currently rejected pool. 

In this section, we show that the probability of selecting one amongst the top $d$ candidates out of a pool of $n$ total candidates, using this strategy with a set of stopping points $(r_{1},r_{2},...,r_{d})$, is given by
\begin{align*}
P(r_{1},r_{2},...,r_{d}) = \sum_{s=1}^{d} \hspace{1ex} \prod_{i=1}^{s-1} \frac{r_{i}-i}{r_{s}-i} \hspace{0.5ex} \cdot P(s,r_{s},r_{s+1}),
\end{align*}
where
\vspace{2ex}

${\textstyle 
\noindent {\textit{$P(s,r,r')$}} = d! \prod_{l=0}^{d-1} \frac{1}{n-l} \hspace{0.5ex} \left( \sum_{k=0}^{s-1} {r-1 \choose k} \hspace{0.5ex} \sum_{i=r}^{r'-1} {n-i \choose d-k-1} \hspace{0.5ex} \prod_{j=k+1}^{s} \frac{r-j}{i-j} \hspace{1ex} + \hspace{1ex} \sum_{k=s}^{d-1} {r-1 \choose k} \sum_{i=1}^{d-k} {r'-r \choose i} \cdot {n-r'+1 \choose d-k-i} \cdot S_{s}(k+i,i) \right)}$
\begin{align*}
     S_{s}(k,m) \hspace{1ex} = \hspace{1ex}1 - \prod_{i=0}^{s-1} \frac{k-m-i}{k-i}.
\end{align*}

For heuristic reasons, we begin this section by first considering the simplified case where $d=2$.
\subsection{The two admissible candidates case: probability computations}

\vspace{2ex}

As a first step towards generalization, we consider the first case with weaker success criteria, where we are required to select one of the best two candidates in the pool, with the same rules as the standard case. In order to maximize the success probability, we naturally extend the standard strategy to two strategies, whose probability distributions will motivate the implementation of a third, adaptive search strategy, which can lead to even better success probabilities.

\subsubsection{First strategy: stopping at the best candidate interviewed so far}
\vspace{2ex}
At first glance, we may want to use the same exact strategy as the standard case, and benefit from the weaker success criteria:

\bigskip

\textbf{Strategy 1:}  \textit{Reject the first $r-1$ candidates, for $r \geq 2$, and then select the first candidate which bests the pool} \\ 
\hspace*{17.5ex} \textit{of rejected candidates.}

\bigskip

\noindent{Defining "success" as the selection of the best or second-to-best candidate, and assuming that both of these candidates are located at positions $n_1$ and $n_2$, there are only three subcases where success is possible: }

\begin{align*}
     & \text{Subcase 1:} \quad r \leq n_1 < n_2 \leq n  \\
     & \text{Subcase 2:} \quad r \leq n_2 < n_1 \leq n  \\
     & \text{Subcase 3:} \quad n_2 < r \leq n_1 \leq n
\end{align*}

We can compute $P(\text{Success } \cap \text{ Subcase 1})$ by considering the probability of success when the two candidates of interest are at some fixed positions $n_1$ and $n_2$, and then letting $n_1$ and $n_2$ iterate through all their possible positions within \textbf{Subcase 1}: 

\begin{align*}
P(\text{Success } \cap  \text{ Subcase 1}) &= \sum_{n_1=r}^{n-1} \hspace{1ex} \sum_{n_2=n_1+1}^{n} P(\text{Success } \cap \text{ Best two candidates at $n_1$ and $n_2$}) \\
    &=\sum_{n_1=r}^{n-1} \hspace{1ex} \sum_{n_2=n_1+1}^{n} P(\text{Best two candidates at $n_1$ and $n_2$}) \\
    &\quad \quad \quad \quad \quad \quad \quad \quad \cdot P(\text{Success } |\text{ Best two candidates at $n_1$ and $n_2$}) \\
    &= \sum_{n_1=r}^{n-1} (n-n_1) \cdot \frac{1}{n} \cdot \frac{1}{n-1} \cdot P(\text{Best candidate before $n_1$ lies in the $r-1$ group}) \\
    &= \frac{1}{n} \cdot \frac{1}{n-1} \sum_{n_1=r}^{n-1} (n-n_1) \cdot \frac{r-1}{n_1-1}.
\end{align*}

\noindent
\noindent

By similar reasoning, we obtain the same expression for $P(\text{Success } \cap \text{ Subcase 2})$. Moreover, because no candidate between $r$ and $n_1$ can best $n_2$, it follows that
\begin{align*}
P(\text{Success } \cap \text{ Subcase 3}) &= P(\text{Subcase 3}) \cdot P(\text{Success } | \text{ Subcase 3}) \\
     &= (r-1) \cdot (n-r+1) \cdot \frac{1}{n} \cdot \frac{1}{n-1}.
\end{align*}

Because success can only occur within these three subcases, the total probability of success for\\ 
\textbf{Strategy 1} is given by
\begin{align*}
    P(r) &= P(\text{Success }  \cap \text{ Subcase 1}) + P(\text{Success } \cap \text{ Subcase 2}) + P(\text{Success } \cap \text{ Subcase 3}) \\
    &= \left( \frac{r-1}{n(n-1)} \right) \left( (n-r+1) + 2\sum_{i=r}^{n-1} \frac{n-i}{i-1}  \right) \\
    &= \left( \frac{r-1}{n} \right) \left(  2\sum_{i=r}^{n}  \frac{1}{i-1} - \left( \frac{n-r+1} {n-1} \right) \right) \\
    &\approx 2 \cdot \left( \frac{r-1}{n} \right) \sum_{i=r}^{n}  \frac{1}{i-1}.
\end{align*}
Notice how this success probability roughly equates twice the one computed for the strict case in the first section.

\vspace{2ex}
    
\subsubsection{Second strategy (satisficing): stopping at best or second-to-best candidates interviewed so far}
\vspace{3ex}

Alternatively, we could also implement the following strategy:

\bigskip

\textbf{Strategy 2:}  \textit{Reject the first $r-1$ candidates, for $r\geq3$, and then select the first candidate which is the best or} \\ 
\hspace*{17.5ex}\textit{second-best relative to them.}

\bigskip
\noindent Using this strategy, there is a further case where success is possible that we need to consider:
\begin{align*}
     \text{Subcase 4}: \quad n_1 < r \leq n_2 \leq n
\end{align*}

$P(\text{Success } \cap \text{ Subcase 1})$ can be computed using a similar method to the one used for \textbf{Strategy 1}:
\newpage
\begin{align*}
P(\text{Success } \cap \text{ Subcase 1}) &= \sum_{n_1=r}^{n-1} \hspace{1ex} \sum_{n_2=n_1+1}^{n} P(\text{Best two candidates at $n_1$ and $n_2$}) \\
&\quad \quad \quad \quad \quad \quad \quad \quad \cdot P(\text{Success } | \text{Best two candidates at $n_1$ and $n_2$}) \\
    &= \sum_{n_1=r}^{n-1} (n-n_1) \cdot \frac{1}{n} \cdot \frac{1}{n-1} \cdot P(\text{Best two candidates before $n_1$ lie in the $r-1$ group}) \\
    &= \frac{1}{n} \cdot \frac{1}{n-1} \sum_{n_1=r}^{n-1} (n-n_1) \cdot \frac{r-1}{n_1-1} \cdot \frac{r-2}{n_1-2}.
\end{align*}

\noindent By similar reasoning, we obtain the same expression for $P(\text{Success } \cap \text{ Subcase 2})$.\\	

Moreover, by the same method:
\begin{align*}
P(\text{Success } \cap \text{ Subcase 3}) &=\sum_{n_2=1}^{r-1} \hspace{1ex} \sum_{n_1=r}^{n} \frac{1}{n} \cdot \frac{1}{n-1}\cdot P(\text{Second-best candidate before $n_1$ lies in the $r-1$ group }|  \\
    &  \hspace{30ex} n_2 \text{ lies in the $r-1$ group}) \\
    &= \frac{1}{n} \cdot \frac{1}{n-1} \cdot (r-1) \sum_{n_1=r}^{n} \frac{r-2}{n_1-2}.
\end{align*}

\noindent And, again, by similar reasoning, we obtain the same expression for $P(\text{Success } \cap \text{ Subcase 4})$.	
\vspace{1ex}

Hence, the total probability of success for \textbf{Strategy 2} is given by 
\begin{align*}
    P(r) &= \left( \frac{2(r-1)(r-2)}{n(n-1)} \right) \left(\sum_{i=r}^{n} \hspace{1ex}  \frac{1}{i-2} + \sum_{i=r}^{n-1} \hspace{1ex}  \frac{n-i}{(i-1)(i-2)}  \right).
\end{align*}

\subsubsection{Comparing the two strategies}

\begin{figure}[htp]
    \centering
    \includegraphics[width=10cm,height=5cm]{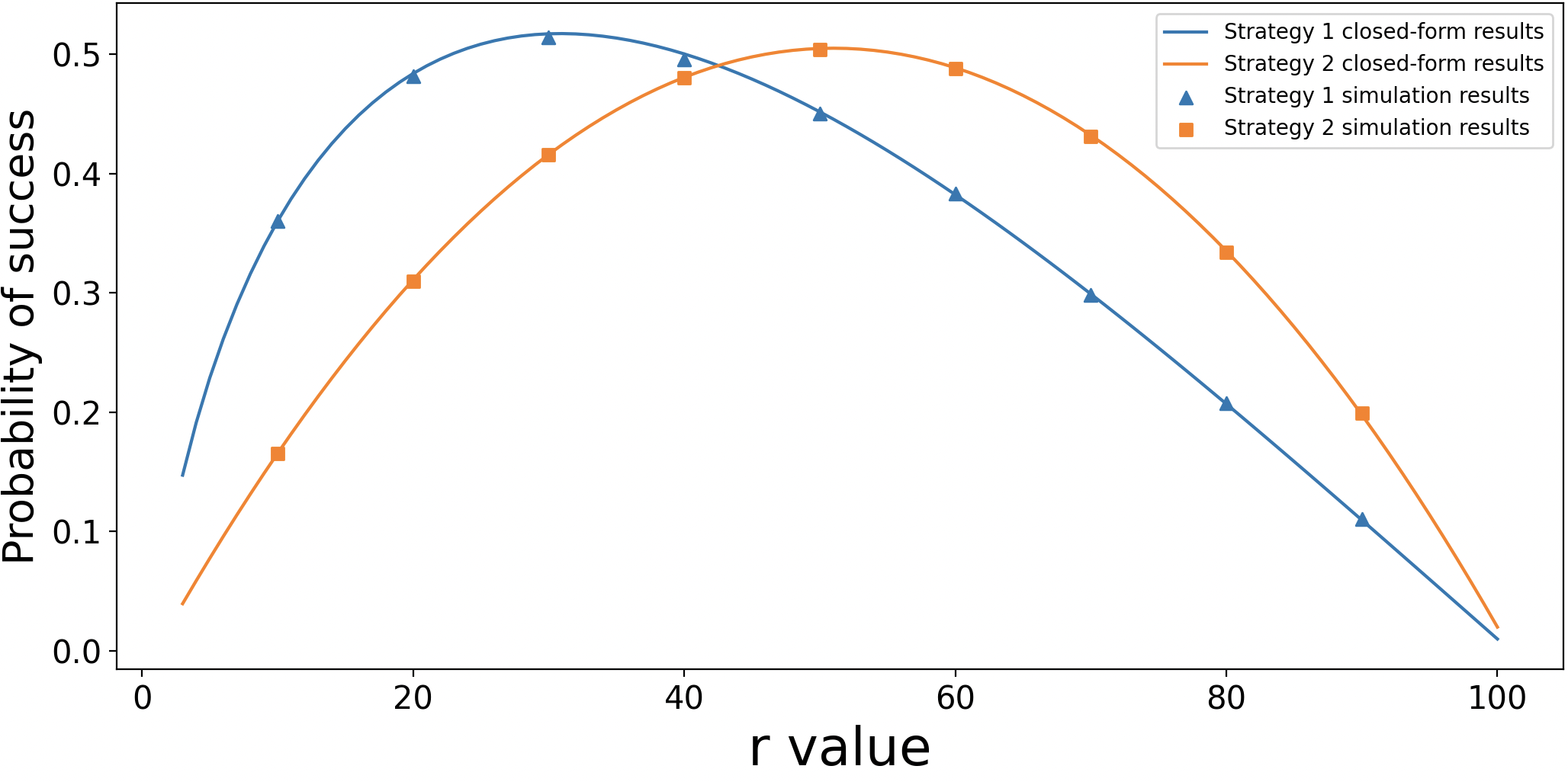}
    \caption{Probability of success for Strategies 1 and 2 for $d = 2$ and $n = 100$. Analytical results (solid curves) agree well with computer simulations (symbols).}
    \label{fig:strat12}
\end{figure}

Using the formulae derived in sections 2.2.1 and 2.2.2, we have graphed the probability of success against the $r$-value for both strategies, and compared them with computer simulations averaged over $100,000$ trials (see the Appendix for more details regarding the code implemented), as shown Figure~\ref{fig:strat12}.

\subsubsection{A third strategy (adaptive search strategy)}
\vspace{2ex}

From the data shown in Figure~\ref{fig:strat12}, it is clear that \textbf{Strategy 1} significantly outperforms \textbf{Strategy 2} at low $r$-values, while the opposite is true at high $r$-values. Hence, this motivates us to merge the two strategies to further optimize their performance:

\bigskip

\textbf{Adaptive search strategy (Strategy 3):}  \textit{Implement \textbf{Strategy 1}, rejecting the first $r_{1}-1$ candidates. If no selection} \\ \hspace*{39ex} \textit{is made by the time you have reached the $(r_{2}-1)^{th}$ candidate, switch to \\ \hspace*{40ex} \textbf{Strategy 2} for the}
 \textit{remaining candidates}.

 \bigskip

The probability of success with this strategy is given by
\begin{align*}
    &P(\text{Success with \textbf{Strategy 1} on first $r_2 - 1$ candidates})\\
    &\quad + P(\text{No selections made with \textbf{Strategy 1} } \cap \text{ Success with \textbf{Strategy 2}}).
\end{align*}

\noindent For the second term, we note:
\begin{align*}
    P(\text{No selections made } &\text{with \textbf{Strategy 1} } \cap \text{ Success with \textbf{Strategy 2}})\\
    &= P(\text{No selections made with \textbf{Strategy 1}}) \cdot P(\text{Success with \textbf{Strategy 2}})\\
    &= P(\text{Best candidate before $r_2$ lies before $r_1$}) \cdot P(\text{Success with \textbf{Strategy 2}})\\
    &= \left(\frac{r_1 - 1}{r_2 - 1}\right) \cdot P(\text{Success with \textbf{Strategy 2}}),
\end{align*}
where to evaluate $P(\text{Success with \textbf{Strategy 2}})$, it suffices to set $r = r_2$ in the formula at the end of section 2.2.2.

\noindent For the first term, we need to derive a slight generalization for the formula in section 2.2.1:
\bigskip

\subsubsection{Strategy 1 on a restricted sample}
\vspace{2ex}
In this section, our aim is to compute the probability of achieving success by \textbf{Strategy 1} on a restricted sample of $r_{2}-1$ candidates, within the total sample of $n$ candidates. Analogously to section 2.2.1, we tackle the problem by first identifying all subcases where success is possible, considering the respective positions $n_1$ and $n_2$ of the best and second-best candidates:

\begin{align*}
     & \text{Subcase 1:} \quad r_1 \leq n_1 < n_2 \leq r_{2}-1  \\
     & \text{Subcase 2:} \quad r_1 \leq n_2 < n_1 \leq r_{2}-1  \\
     & \text{Subcase 3:} \quad n_2 < r_1 \leq n_1 \leq r_{2}-1 \\
     & \text{Subcase 4:} \quad r_1 \leq n_2 \leq r_{2}-1 < n_1  \\
     & \text{Subcase 5:} \quad r_1 \leq n_1 \leq r_{2}-1 < n_2  \\
\end{align*}

\noindent For the first two subcases, it suffices to repeat a similar proof to the one for \textbf{Subcase 1} and \textbf{Subcase 2} of section 2.2.1, with restricted range for $n_1$ and $n_2$:	 
\begin{align*}
P(\text{Success } \cap \text{ Subcase 1 or 2}) &= 2 \sum_{n_1=r_1}^{r_{2 }-2} \hspace{1ex} \sum_{n_2=n_1+1}^{r_{2 }-1} P(\text{Success } \cap \text{ Best two candidates at $n_1$ and $n_2$}) \\
    &= \frac{2}{n} \cdot \frac{1}{n-1} \sum_{n_1=r_1}^{r_{2 }-2} \hspace{1ex} (r_{2 }-1-n_1) \cdot \frac{r_{1}-1}{n_1-1}.
\end{align*}

\noindent Similarly, $P(\text{Success } \cap \text{ Subcase 3}) = \frac{1}{n} \cdot \frac{1}{n-1} \cdot (r_{1}-1) \cdot (r_{2}-r_{1})$, and, by similar reasoning:
\begin{align*}
P(\text{Success } \cap \text{ Subcase 4 or 5}) = \frac{2}{n} \cdot \frac{1}{n-1} \cdot (n-r_{2}+1) \sum_{n_1=r_1}^{r_{2 }-1}  \frac{r_{1}-1}{n_1-1}.
\end{align*}

\noindent Hence, after simplifying:
\begin{align*}
    P(\text{Success with \textbf{Strategy 1} on first $r_2 - 1$ candidates}) &= \left( \frac{r_{1}-1}{n(n-1)} \right) \left( (r_{2}-r_{1}) + 2\sum_{i=r_{1}}^{r_{2}-1} \hspace{1ex}  \frac{n-i}{i-1}  \right).
\end{align*}

\subsubsection{Strategy 3 probability computation}
\vspace{2ex}

We then compute the total probability of success $P(r_{1},r_{2})$ for \textbf{Strategy 3}:
\begin{align*}
P(r_1, r_2) &= \left( \frac{r_{1}-1}{n(n-1)} \right) \left( 2(r_{2}-2) \left(\sum_{i=r_2}^{n} \hspace{1ex}  \frac{1}{i-2} + \sum_{i=r_2}^{n-1} \hspace{1ex}  \frac{n-i}{(i-1)(i-2)} \right) + (r_{2}-r_{1}) + 2\sum_{i=r_{1}}^{r_{2}-1} \hspace{1ex}  \frac{n-i}{i-1} \right) \\
     &= \left( \frac{r_{1}-1}{n(n-1)} \right) \left( 2(r_{2}-2) \left(\sum_{i=r_2}^{n} \hspace{1ex}  \frac{n-1}{(i-1)(i-2)} \right) + (r_{2}-r_{1}) + 2 \sum_{i=r_{1}}^{r_{2}-1} \left( \hspace{1ex}  \frac{n-1}{i-1} -1 \right) \right)\\
      &= \left( \frac{r_{1}-1}{n(n-1)} \right) \left( 2(r_{2}-2)(n-1) \left(\sum_{i=r_2}^{n} \hspace{1ex}  \frac{1}{i-2} - \sum_{i=r_2}^{n} \hspace{1ex}  \frac{1}{i-1} \right) + (r_{1}-r_{2}) + 2 \sum_{i=r_{1}}^{r_{2}-1} \left( \hspace{1ex}  \frac{n-1}{i-1}\right)  \right) \\
      &= \left( \frac{2(r_{1}-1)}{n(n-1)} \right) \left( (n-r_{2}+1) + \frac{(r_{1}-r_{2})}{2} + \sum_{i=r_{1}}^{r_{2}-1} \left( \hspace{1ex}  \frac{n-1}{i-1}\right)  \right).
\end{align*}

\vspace{2ex}

In Figure~\ref{fig:strat3}, we plot a heat map illustrating the probability of success against $(r_{1},r_{2})$ values for this search adaptive strategy, with contour lines showing the maximum values achieved by the two strategies discussed so far:
\begin{figure}[htp]
    \centering
    \includegraphics[width=8cm,height=6.5cm]{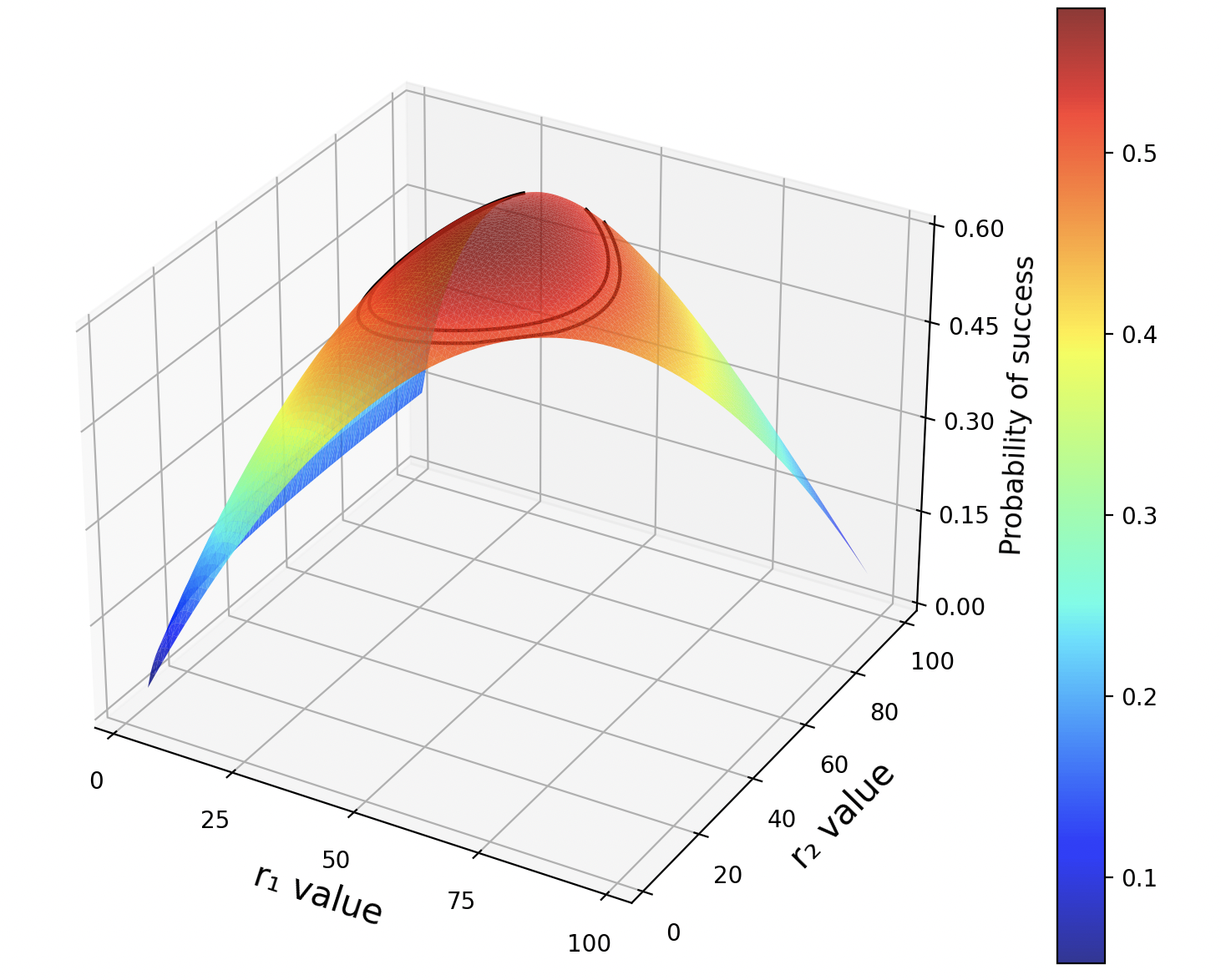}
    \caption{Probability of success as a function of stopping thresholds $(r_1, r_2)$ by using an adaptive search strategy,  for $d = 2$ and $n = 100$. The countour lines correspond to the maximum probability of success given by the prior Strategies 1 and 2.}
    \label{fig:strat3}
\end{figure}

Table 1 compares the corresponding results obtained through the formula derived in section 2.2.6 with computer simulations averaged over 100,000 trials (see the Appendix for more details regarding the code implemented):
\begin{center}
\begin{tabular}{ |c|c|c| } 
\hline
$(r_{1},r_{2})$ & $Success \hspace{0.5ex} probability \hspace{0.5ex} by \hspace{0.5ex} closed \hspace{0.5ex} form \hspace{0.5ex} solutions \hspace{0.5ex} (Strategy 3)$ & $Success \hspace{0.5ex} probability \hspace{0.5ex} by \hspace{0.5ex} simulations \hspace{0.5ex} (Strategy 3)$  \\
\hline
$(20,50)$ & $0.50439$ & $0.50497$ \\  
$(20,60)$ & $0.51804$ & $0.52063$ \\ 
$(20,70)$ & $0.52043$ & $0.51970$ \\  
$(30,50)$ & $0.54855$ & $0.54934$ \\ 
$(30,60)$ & $0.56939$ & $0.56910$ \\  
$(30,70)$ & $0.57304$ & $0.57300$ \\ 
$(40,50)$ & $0.54252$ & $0.54046$ \\ 
$(40,60)$ & $0.57056$ & $0.57028$ \\ 
$(40,70)$ & $0.57546$ & $0.57539$ \\ 
\hline
\label{tab11}
\end{tabular}
\captionof{table}{Comparison of closed-form success probability and simulations}
\end{center}
\vspace{2ex}

\subsection{The $d$ admissible candidates case: probability computations}
\vspace{2ex}
In this section, we aim to generalize the results of the previous section, establishing a formula governing the total probability of success $P(r_{1},r_{2},...,r_{d})$ when required to select one of the best \textit{d} candidates within the pool of \textit{n} candidates. Analogously to section 2.2, we will first determine a formula for the total probability of success for each of the \textit{d} standard strategies within a restricted sample of the candidates pool in a stepwise manner. This will then allow their combination into a final strategy maximizing the overall probability of success based on mutually exclusive events.

\subsubsection{General standard strategy on a restricted sample of candidates}
\vspace{1ex}
When required to select one amongst the best \textit{d} candidates, there are \textit{d} possible strategies one could possibly implement. Hence, for each positive integer-valued $2 \leq s \leq d$, we can define:

\bigskip

\textbf{Strategy \textit{s}:}  \textit{Reject the first $r-1$ candidates and then select the first candidate which ranks at least \textit{s}th} \\ \hspace*{14ex} \textit{relative to the pool of rejected candidates.}

\bigskip

The aim of this subsection is to determine, for each $s \leq d$, the total probability of success $P(s,r,r')$ when implementing Strategy \textit{s} on a restricted sample of $r'-1   \leq n$ candidates, rejecting the first ${r-1}$ candidates.

\bigskip

Assuming that the best \textit{d} candidates, ignoring rank, hold positions $n_{1}<n_{2}<...<n_{d} \leq n$, it is convenient to categorize the subcases where success is possible according to the number of the best \textit{d} candidates in the rejected pool, leading us to consider the following \textit{d} general subcases: 

\begin{align*}
     & \text{Subcase 0:} \quad r \leq n_1 < n_2 <...< n_d \\
     & \text{Subcase 1:} \quad n_1 < r \leq n_2 < n_3 <...< n_d  \\
     & \text{Subcase 2:} \quad n_1 < n_2 < r \leq n_3 < n_4 <...< n_d  \\
     & ... \\
     & \text{Subcase \textit{k}:} \quad n_1 < n_2 <...< n_k < r \leq n_{k+1} <...< n_d \\ 
     & ... \\
     & \text{Subcase \textit{d-1}:} \quad n_1 < n_2 <...< n_{d-1} < r \leq n_{d}
\end{align*}

Before proceeding, it is helpful to define two pertinent expressions:

\subsubsection{\textbf{Permutations of consecutive candidates}}
\vspace{1ex}

Given some $n,x \in \mathbb{Z}^+$, where $x\geq n$, notice that
\begin{align*}
 & \frac{1}{n!} \prod_{i=0}^{n-1}(x-i) = {x \choose n}.
\end{align*}

\noindent{Then, the total permutations of $k \in \mathbb{Z}^+$ consecutive candidates between positions \textit{a} and \textit{b}, where the \textit{i}th candidate holds position $n_{k-i+1}$ and $a \leq n_{k} < n_{k-1} <...< n_{1} \leq b$, are given by} \\
\begin{align*}
     & \sum_{n_{k}=a}^{b-k+1} \sum_{n_{k-1}=n_{k}+1}^{b-k+2} ... \sum_{n_{i}=n_{i+1}+1}^{b-k+i} ... \sum_{n_{1}=n_{2}+1}^{b} \cdot \hspace{1ex} (1) \hspace{1ex} = \hspace{1ex} \frac{1}{k!} \prod_{i=-1}^{k-2}(b-a-i) \hspace{1ex} = \hspace{1ex} \frac{1}{k!} \prod_{i=0}^{k-1}\left((b-a+1)-i\right) \hspace{1ex} = \hspace{1ex} {{b-a+1} \choose k}.
\end{align*}

\vspace{3ex}
\noindent{The equality above can be shown by induction on $k$ and $b$ respectively, for $b \geq (a+k-1$):} \\
\begin{align*}
     & \sum_{n_{k}=a}^{b-k+1} \sum_{n_{k-1}=n_{k}+1}^{b-k+2} ... \sum_{n_{i}=n_{i+1}+1}^{b-k+i} ... \sum_{n_{1}=n_{2}+1}^{b} \cdot \hspace{1ex} (1) \hspace{1ex}
      = \hspace{1ex} \sum_{n_{k}=a}^{b-k+1} 
     \left(   \frac{1}{(k-1)!} \prod_{i=-1}^{k-3}(b-(n_{k}+1)-i)  \right) \hspace{0.5ex} \left[induction \hspace{0.5ex} hypothesis \hspace{0.5ex} on \hspace{0.5ex} (k-1) \right] \\ 
     & = \frac {1}{(k-1)!}  \prod_{i=-1}^{k-3}(b-a-i-1) \hspace{0.5ex} + \hspace{0.5ex} \frac {1}{(k-1)!} \sum_{n_{k}=a}^{(b-1)-k+1} \prod_{i=-1}^{k-3}((b-1)-(n_{k}+1)-i)       \\
     & = \frac {1}{(k-1)!}  \prod_{i=-1}^{k-3}(b-a-i-1) \hspace{0.5ex} + \hspace{0.5ex} \frac {1}{k!} \prod_{i=-1}^{k-2}((b-1)-a-i) \hspace{2ex} \left[induction \hspace{0.5ex} hypothesis \hspace{0.5ex} on \hspace{0.5ex} (b-1) \right] \\ 
     & = \frac {1}{k!} \left( ((b-1)-a+2)  \prod_{i=-1}^{k-3}((b-1)-a-i)     \right) \\
     & = \frac {1}{k!}  \prod_{i=-1}^{k-2}(b-a-i). \hspace{2ex} \blacksquare
\end{align*}

\vspace{1ex}
\subsubsection{Selections}
\vspace{1ex}

Given some $k \in \mathbb{Z}^+$, suppose that \textit{k} candidates hold positions strictly below \textit{r'}, and that \textit{m} of these hold positions greater than or equal to \textit{r}, where $m \leq k \leq r'-1$. In this subsubsection, we compute the probability $S_{s}(k,m)$ that, given some $s \leq k-m$, at least one of the \textit{m} candidates ranks at least \textit{s}th relative to the \textit{k} candidates.

Notice that this doesn's occur if and only if the best $s$ candidates all lie within the $k-m$ positions strictly below \textit{r}, where the remaining $k-s$ candidates can be arranged without restrictions. Hence, by counting permutations:

\begin{align*}
     & 1 - S_{s}(k,m) \hspace{1ex} = \hspace{1ex} \frac{\frac{(k-m)!}{(k-m-s)!} \cdot (k-s)!}{k!},
\end{align*}

so that

\begin{align*}
     & S_{s}(k,m) \hspace{1ex} = \hspace{1ex}1 - \prod_{i=0}^{s-1} \frac{k-m-i}{k-i}.
\end{align*}

\bigskip

\vspace{2ex}

\noindent{Having defined the expressions above, we will now proceed towards computing $P(s,r,r')$ by partitioning our total $d-1$ subcases into two types (defined in terms of the general Subcase \textit{k}), and compute their respective probabilities in the following two subsubsections.}

\subsubsection{Subcases where $0 \leq k \leq s-1$}
\vspace{1ex}
Note that, for this type of subcase, the probability of success is independent of the specific ranks of the best \textit{d} candidates (since the strategy implemented guarantees the selection of any of the best \textit{d} candidates whenever this is interviewed), but is dependent on the ranks of candidates before the $n_{k+1}$th position (since \textbf{Strategy s} does not require a candidate to be amongst the best \textit{d} candidates for selection). 

Thus, using an analogous approach to Section 2.2.1:\\ 

$P(\text{Success } \cap \hspace{0.5ex}  \text{Subcase} \hspace{1ex} k \leq s-1)$
\begin{align*}
&= \hspace{1ex} \sum_{n_1=1}^{r-k} \hspace{1ex} \sum_{n_2=n_1+1}^{r-k+1} ... \sum_{n_k=n_{k-1}+1}^{r-1} \hspace{1ex} \sum_{n_{k+1}=r}^{r'-1} \hspace{1ex} \sum_{n_{k+2}=n_{k+1}+1}^{n-(d-(k+2))}...\hspace{1ex} \sum_{n_{d}=n_{d-1}+1}^{n}    P(\text{Success } \cap \text{Best} \hspace{1ex} d \hspace{1ex} \text{candidates at $n_1,...,n_d$})\\
&= \hspace{1ex} {r-1 \choose k} \cdot \sum_{n_{k+1}=r}^{r'-1} {n-n_{k+1} \choose d-k-1} \cdot P(\text{Best} \hspace{1ex} d \hspace{1ex} \text{candidates at $n_1,...,n_d$}) \cdot P(\text{Success } | \hspace{1ex} \text{Best} \hspace{1ex} d \hspace{1ex} \text{candidates at $n_1,...,n_d$}) \\
&= \hspace{1ex} {r-1 \choose k} \cdot d! \prod_{i=0}^{d-1} \frac{1}{n-i} \hspace{0.5ex} \cdot \sum_{n_{k+1}=r}^{r'-1} {n-n_{k+1} \choose d-k-1} \hspace{0.5ex} \\
& \hspace{24ex} \cdot P(\text{Best $s$ candidates before $n_{k+1}$ in the $r-1$ group } | \hspace{1ex} \text{Best} \hspace{1ex} d \hspace{1ex} \text{candidates at $n_1,...,n_d$}) \\
&= \hspace{1ex} {r-1 \choose k} \cdot d! \prod_{i=0}^{d-1} \frac{1}{n-i} \hspace{0.5ex} \cdot \sum_{n_{k+1}=r}^{r'-1} {n-n_{k+1} \choose d-k-1} \cdot \hspace{0.5ex} \prod_{j=k+1}^{s} \frac{r-j}{n_{k+1}-j}.
\end{align*}

\subsubsection{Subcases where $s \leq k \leq d-1 $}
\vspace{1ex}

Conversely to the previous subsubsection, notice that for this type of subcase the probability of success is dependent on the specific ranks of the best \textit{d} candidates (since the more highly selective strategy may reject even such candidates unless they rank sufficiently high relative to the rejected pool), but independent of the ranks of candidates before the $n_{k+1}$th position (since \textbf{Strategy s}, in this case, requires a candidate to rank higher than at least one of the best \textit{k} rejected candidates for selection, hence to be one of the best \textit{d} candidates overall).

Therefore, due to the rank-dependence of the best \textit{d} candidates, we will need a slightly different approach:

\vspace{3ex}
$P(\text{Success } \cap \hspace{0.5ex}  \text{Subcase} \hspace{1ex} k \geq s)$
\vspace{-1ex}
\begin{align*}
&= \hspace{1ex} \sum_{i=1}^{d-k} \hspace{1ex} P\left((\text{Success } \cap \hspace{0.5ex}  \text{Subcase} \hspace{1ex} k \geq s) \hspace{0.5ex} \cap \hspace{0.5ex} n_{1} <...< n_{k} < r \leq n_{k+1} < ...< n_{k+i} < r'\leq n_{k+i+1} <...<n_{d} \leq n  \right) \\
&= \hspace{1ex} \sum_{i=1}^{d-k} \hspace{1ex} P(n_{1} <...< n_{k} < r \leq n_{k+1} < ...< n_{k+i} < r'\leq n_{k+i+1} <...<n_{d} \leq n) \\
& \hspace{16ex} \cdot P\left((\text{Success } \cap \hspace{0.5ex}  \text{Subcase} \hspace{1ex} k \geq s) \hspace{0.5ex} | \hspace{0.5ex} n_{1} <...< n_{k} < r \leq n_{k+1} < ...< n_{k+i} < r'\leq n_{k+i+1} <...<n_{d} \leq n  \right)  
\end{align*}
\begin{align*}
&= \hspace{1ex} \sum_{i=1}^{d-k} \frac{{r-1 \choose k} \cdot {r'-r \choose i} \cdot {n-r'+1 \choose d-k-i}}{{n \choose d}} \cdot P(\text{at least one candidate in } n_{k+1},...,n_{k+i} \text{ ranks at least } s \text{th relative to candidates in } n_{1},...,n_{k+i})
\end{align*}
\begin{align*}
\hspace{-32ex}= \hspace{1ex} {r-1 \choose k} \cdot d! \prod_{j=0}^{d-1} \frac{1}{n-j} \hspace{0.5ex} \cdot \sum_{i=1}^{d-k} {r'-r \choose i} \cdot {n-r'+1 \choose d-k-i} \cdot S_{s}(k+i,i).
\end{align*}
\bigskip

\subsubsection{Computing the probability of success for any given Strategy $s$ on a restricted sample of candidates}
\vspace{1ex}

Using the results of the two previous subsubsections, and the fact that success can only occur within the $d$ subcases listed in section 2.3.1, we can finally compute the probability of success for any given Strategy \textit{s} on a restricted sample of candidates between $r, \hspace{0.5ex} r'-1$:

\begin{align*}
\noindent {\textit{$P(s,r,r')$}} &= \sum_{k=0}^{d-1} \hspace{1ex} P(\text{Success } \cap \hspace{0.5ex}  \text{Subcase} \hspace{1ex} k) \\
&= \sum_{k=0}^{s-1} \hspace{1ex} P(\text{Success } \cap \hspace{0.5ex}  \text{Subcase} \hspace{1ex} k \leq s-1) \hspace{1ex} + \hspace{1ex} \sum_{k=s}^{d-1} \hspace{1ex} P(\text{Success } \cap \hspace{0.5ex}  \text{Subcase} \hspace{1ex} k \geq s)\\
&= d! \prod_{l=0}^{d-1} \frac{1}{n-l} \hspace{0.5ex} \cdot \\
&\hspace{0ex} \left( \sum_{k=0}^{s-1} {r-1 \choose k} \hspace{0.5ex} \sum_{i=r}^{r'-1} {n-i \choose d-k-1} \hspace{0.5ex} \prod_{j=k+1}^{s} \frac{r-j}{i-j} \hspace{1ex} + \hspace{1ex} \sum_{k=s}^{d-1} {r-1 \choose k} \sum_{i=1}^{d-k} {r'-r \choose i} \cdot {n-r'+1 \choose d-k-i} \cdot S_{s}(k+i,i) \right).
\end{align*}

\subsubsection{General adaptive search strategy}
\vspace{1ex}

Following the motivation of Section 2.2.4, we can now generally define:

\bigskip

\hspace{-3ex}\textbf{Strategy \textit{d+1}:}  \textit{Reject the first $r_{1}-1$ candidates. For each $s \leq d$, until no selection is made, implement }\\ \hspace*{17ex} \textit{ \textbf{Strategy s} between the $r_{s}^{th}$ and the $(r_{s+1}-1)^{th}$ candidate, where $r_{d+1}=n+1$.}

\bigskip

Armed with the results of the previous subsection, we can now compute the total probability of success for this adaptive search strategy. Note that, for $s=1$, the product operator below evaluates to $1$, giving the probability of success for \textbf{Strategy 1} between $r_{1},r_{2}$, as desired.

\begin{align*}
P(r_{1},r_{2},...,r_{d}) &= \sum_{s=1}^{d} \hspace{1ex} \prod_{i=1}^{s-1} P(\text{Strategy } i \text{ made no selections}) \hspace{0.5ex} \cdot P(s,r_{s},r_{s+1}) \\
&= \sum_{s=1}^{d} \hspace{1ex} \prod_{i=1}^{s-1} P(\text{best } i \text{ candidates before position } r_{i+1}-1 \text{ in the first } r_{i}-1 \text{ positions)}  \hspace{0.5ex} \cdot P(s,r_{s},r_{s+1}) \\
&= \sum_{s=1}^{d} \hspace{1ex} \prod_{i=1}^{s-1} \hspace{0.5ex} \prod_{j=1}^{i} \frac{r_{i}-j}{r_{i+1}-j} \hspace{0.5ex} \cdot P(s,r_{s},r_{s+1}) \\
&= \sum_{s=1}^{d} \hspace{1ex} \prod_{i=1}^{s-1} \frac{r_{i}-i}{r_{s}-i} \hspace{0.5ex} \cdot P(s,r_{s},r_{s+1}),
\end{align*}

where the final step can be shown by induction (the statement holds trivially for $s=1$):

\begin{align*}
\prod_{i=1}^{s-1} \hspace{0.5ex} \prod_{j=1}^{i} \frac{r_{i}-j}{r_{i+1}-j} &= \left(\prod_{j=1}^{s-1} \frac{r_{s-1}-j}{r_{s}-j}  \right) \cdot\left( \prod_{i=1}^{s-2} \hspace{0.5ex} \prod_{j=1}^{i} \frac{r_{i}-j}{r_{i+1}-j} \right) \\
&= \left( \frac{r_{s-1}-(s-1)}{r_{s}-(s-1)} \cdot \prod_{j=1}^{s-2} \frac{r_{s-1}-j}{r_{s}-j}  \right) \cdot \left( \prod_{i=1}^{s-2}  \frac{r_{i}-i}{r_{s-1}-i} \right) \\
&= \frac{r_{s-1}-(s-1)}{r_{s}-(s-1)} \cdot \prod_{i=1}^{s-2}  \frac{r_{i}-i}{r_{s}-i} \hspace{1ex} = \hspace{1ex} \prod_{i=1}^{s-1} \frac{r_{i}-i}{r_{s}-i}.\\
\end{align*}

Hence, for each added threshold $r_{s}$, this strategy bumps up the overall success probability monotonically with respect to $d$. Though not surprising, it ensures a greater probability of finding sub-optimal choices above the selection criteria. 

\section{Computational results on optimization of search success}

In this section, we verify the results derived in the previous section using computer simulations, and propose an algorithm to maximize success probability within a reasonable computation time.

\subsection{Simulating the outcomes of an arbitrary number of trials}

\vspace{2ex}

The algorithm below simulates the problem $numtrials$ times for finding the best $d$ candidates, over $n$ total candidates, given a list of selection thresholds $(r_{1},r_{2},...,r_{d})$. It can be used to verify the analytical results of the formula derived in section 2.3.7 by using simulations. The full Python code for the algorithm can be found in the Appendix.

\SetKwInOut{Parameter}{Parameters}
\begin{algorithm}[!htb]
\SetAlgoLined
\Parameter{$n$, $d$, $numtrials$,  selection thresholds $(r_{1},r_{2},...,r_{d})$, where $r_{d+1}=n+1$}

 \For{$trial = 1, \cdots, numtrials$}{

  Generate list of $n$ candidates with random ranking orders;

  Set $leastAcceptable$ to $d$th highest ranking in the list;

  Set selection round $s$ to $1$;

  \For{$s = 1 \hspace{0.5ex} to \hspace{0.5ex} d$} { 

  \For{$i = r_{s} \hspace{0.5ex} to \hspace{0.5ex} r_{s+1}-1$} { 

  \If{ $i$th candidate has greater ranking than $s$th best candidate so far}
   {\eIf{ $i$th candidate has greater ranking than $leastAcceptable$} {Record successful trial}
     {Record failed trial}}
     
  }

  }

  Record failed trial

 }

Compute percentage of successful trials out of $numtrials$ total trials.

 \caption{  \textbf{Simulating the selection process}}
\end{algorithm}

For instance, we simulate the problem  one million times for the best $5$ candidates, over $100$ total candidates, with selection thresholds $50,60,65,75,80$ (see Appendix for more details on the specific code implemented).

In this case, the final formula derived in section 2.3.7 computes a success probability of $0.829427$ (6 d.p.), and further similar tests show excellent agreement with probability discrepancies generally below $10^{-3}$.

\bigskip
\bigskip

\subsection{Maximizing success probability: the brute force method}
\vspace{2ex}
In order to determine the values for $r_{1},r_{2},...,r_{d}$ yielding the maximum success probability, we can use the following direct procedure. The full Mathematica code is provided in the Appendix.

\bigskip

1) Generate a list $rlist[d,n]$ containing all the possible combinations of $r_{1},r_{2},...,r_{d}$ for given values of $d,n$.

\vspace{1ex}

2) Use the formula derived at the end of Section 2 to compute the probability of success for given $d,n$ values, 
\hspace*{6ex} and a list $rlist$ containing the current choices of $r_{1},r_{2},...,r_{d}$.

\vspace{1ex}

3) Compute the maximum probability and the optimal choices of $r_{1},r_{2},...,r_{d}$ for given values of $d,n$ by \\
\hspace*{6ex} exhaustively iterating through  $rlist[d,n]$.

\vspace{3ex}

Despite that this method is guaranteed to find the optimal solution to the problem, it is very inefficient due to the high computational run-times. In fact, as the formula below shows, the total number of lists $L(d,n)$ the algorithm is required to iterate through ramps up extremely quickly as values of $d,n$ increase:

\begin{align*}
    L(d,n) \hspace{0.5ex}= \hspace{0.5ex} \frac{1}{d!} \prod_{i=1}^{d} \hspace{0.5ex}(n-i) \hspace{0.5ex}= \hspace{0.5ex} \frac{n-d}{n} {n \choose d}.
\end{align*}

For instance, if we wanted to select one amongst the best $d=6$ candidates, out of a total of $n=100$ candidates, we would already need to iterate over above one billion of possible combinations for $r_{1},r_{2},...,r_{6}$, making the algorithm very inconvenient on a practical level.

This issue motivates the method of simulated annealing in the following section.

\bigskip

\subsection{Maximizing success probability: Simulated annealing}

\vspace{2ex}

This method aims to find the greatest possible success probability for given $d,n$ values within an arbitrary number $numIterations$ of searches in the total search space $rlist[d,n]$.

\bigskip

We begin at an initial state $rlist0[d,n]$ (a best guess for the optimal choices of $r_{1},r_{2},...,r_{d}$) and then, given a current state \textit{rlistCurrent}, the acceptance probability function $P$ determines the probability of moving to a random neighboring state \textit{rlistCandidate}, based on the relative difference in the states' success probabilities and the annealing schedule temperature $T$. The probability of success $p$ for a given state is determined by the formula derived at the end of Section 2 (and computed in the exact same way as in Step 2 of Section 3.2). $T$ depends on the number of current iterations $i$ relative to the total iterations $numIterations$ allowed, and a \textit{RescaleFactor} needed to amplify the relatively small probability discrepancies between neighboring states. In general, as the algorithm approaches the maximum number of iterations, $T$ gradually approaches zero, reducing the probability of a change of state.

\bigskip

The pseudocode below illustrates the algorithm. The full Mathematica code implementing the algorithm, our full definitions for $rlist0[d,n]$, $T$, and $P$, and the neighbour-generating procedure are shown in the Appendix.

\SetKwInOut{Parameter}{Parameters}

\begin{algorithm}[!htb]

\SetAlgoLined
\Parameter   {$d$, $n$, $numIterations$}
\SetAlgorithmName{Algorithm}

 Set current state $s \leftarrow rlist0[d,n]$;
 \vspace{0.5ex}

 \For{$i = 1$ \text{to} $numIterations$}{
 \vspace{1ex}
  Set temperature $T \leftarrow  (\frac{n-i}{n\cdot RescaleFactor}) $;
  \vspace{1.5ex}
  
  Pick a random neighbour $s_{new}$ ;
   \vspace{1.5ex}
   
  Set acceptance probability $P(s,s_{new}) \leftarrow \frac {1}{1+e^{\frac{p(s)-p(s_{new})}{T}}}$;
  \vspace{1ex}
  
  \If {$P(s,s_{new}) > random(0,1) \footnotemark$} {\vspace{1ex} Set $s \leftarrow s_{new}$;}
 }
 \vspace{1ex}
 Output final state $s$;
 
 \caption{Simulated annealing for optimal solutions}
\end{algorithm}
\footnotetext{This function call generates a random number uniformly distributed over (0,1).}

Despite that simulated annealing is not guaranteed to find the optimal solution to the problem, it can find extremely close estimates by sampling only a tiny fraction of the total search space.

Table 2 illustrates the efficiency of this method by comparing the results it produces within a negligible computation time, involving merely $100$ iterations, to the maximum probabilities computed by brute force over a considerably longer period (which is proportional to the formula for $L(d,n)$ shown above):

\begin{center}
\begin{tabular}{ |c|c|c||c|c|c| } 
\hline
$Parameters \hspace{0.5ex} (d,n)$ & $Simulated \hspace{0.5ex} annealing$ & $optimal \hspace{0.5ex} r_{i} \hspace{0.5ex} values$ & $Brute \hspace{0.5ex} force$ & $optimal \hspace{0.5ex} r_{i} \hspace{0.5ex} values$ & Discrepancy \\ 
\hline
$2,20$ & $0.604616$ &$(8,14)$& $0.604616$ &$(8,14)$& $0.000000$ \\ 
$3,20$ & $0.747476$ &$(7,13,16)$& $0.747476$ &$(7,13,16)$& $0.000000$ \\ 
$4,20$ & $0.842694$ &$(7,12,15,17)$& $0.842694$ &$(7,12,15,17)$& $0.000000$ \\ 
$2,30$ & $0.594123$ &$(11,21)$& $0.594123$ &$(11,21)$& $0.000000$ \\ 
$3,30$ & $0.734922$ &$(11,18,24)$& $0.734922$ &$(11,18,24)$& $0.000000$ \\ 
$2,50$ & $0.585779$ &$(18,34)$& $0.585779$ &$(18,34)$& $0.000000$ \\ 
$2,100$ & $0.579554$ &$(36,68)$& $0.579561$ &$(35,67)$& $0.000007$ \\ 
\hline
\end{tabular}
\captionof{table}{Comparison of simulated annealing vs. direct brute force method}
\end{center}

Although the runtime complexity of the brute force method does not currently allow to verify whether deviations from the maximum probability are as promising for higher $d,n$ values, these initial results demonstrate how efficient simulated annealing can be at rapidly finding optimal solutions when equipped with an appropriate selection of parameters.

Though not surprising, it is worth noting that the probability of success for a fixed $n$ monotonically increases when increasing $d$. This concept was illustrated analytically in section 2.2.1 where it has been shown that, for a fixed $n$ and single stopping point $r$, the probability of success almost doubles for $d=2$, when compared to the standard case where $d=1$. Similarly, for any $d > 1$, it is sufficient to apply the strict selection criteria of strategy 1 (giving inferior success probabilities than the optimal adaptive search strategy) to observe that we are guaranteed a higher success probability than the standard case, since we would have a more relaxed success criteria under identical selection criteria.

\section{Discussion and conclusion}

The present work centers on the extension of the classical secretary problem by considering suboptimal choices as successful hires, as well as the use of an adaptive search strategy with multiple stopping criteria that allow closed-form calculation of the probability of hiring any of the top $d$ candidates. Our approach provides a more comprehensive and practical solution to the original secretary problem, as we take into account real-world scenarios where hiring the best candidate may not always be possible. Furthermore, applying the simulated annealing method as an alternative computation tool offers a faster and more effective optimization as compared to the brute force exhaustive search method. Our results have implications for various fields, where decision-making under uncertainty is a common challenge~\cite{abdel1982secretary,seale2000optimal,Smith1}, including personnel selection and resource allocation. 

In order to further develop the solution to the problem presented in this paper, a crucial step is to determine whether it is possible to derive algebraically a closed expression directly evaluating the optimal selection thresholds, hence maximum success probability, for arbitrary values of $d,n$. The derivation of such formula would drastically simplify the computational aspect of the problem introduced in section 3, and potentially provide a more complete understanding of its solution. However, the presence of an arbitrary number of variables (due to the weaker selection criteria) and the complexity of the general probability formula computed in section 2.3.7 have made it difficult to follow an approach relatable to the one used for the standard case to find optimal strategies~\cite{kennedy1995particle}. 

Moreover, a related secondary aspect involves carrying out a more rigorous verification of the effectiveness of the probability maximizing method proposed in section 3.3. This can either be done by solving the more fundamental issue described above, providing a method to directly compute optimal selection thresholds and maximum success probability for arbitrary parameters, or, alternatively, by using greater computation power to evaluate discrepancies in less trivial cases. If such verification finds the current method to be unsuitable at greater values of $d,n$, the latter can be improved by optimizing the temperature function of simulated annealing (e.g. through \textit{RescaleFactor} as shown in Algorithm 2) and the total number of iterations. Alternatively, we could adopt particle swarm optimization to enhance the coverage of the search space~\cite{kennedy1995particle}. 

Inspired by the comparison of Strategies 1 and 2 presented in Section 2, our present work investigates an adaptive search strategy with arbitrarily many $d$ stropping criteria. Previously, Gusein-Zade~\cite{Gusein-Zade} and Gilbert and Mosteller~\cite{Gilbert}, respectively, have explored the base case where $d=2$ and computed its limit probability of success as $n\to\infty$ to $\approx0.574$ as well as asymptomatic behavior of general $d$ values. More importantly, we are able to derive closed-form solutions for the probability of success in securing any of the top $d$ candidates. Unlike in Ref.~\cite{Bearden}, our work does not give preference to any candidate within the top $d$ candidates over another: choosing the best candidate, 2nd best candidate, or $d$th best candidate are all seen as equal successes, and not choosing a candidate within the top $d$, no matter their final rank, is seen as an equal failure. Moreover, unlike in Ref.~\cite{Wu}, our work treats all the candidates, including the best $d$, as directly comparable items.

One promising extension for future work is to incorporate game theory~\cite{perc2010coevolutionary} into the secretary problem to account for multiple decision makers~\cite{chen1997secretary} having idiosyncratic preferences within the selection committee~\cite{alpern2017secretary}. Extension like this will help inform fair and efficient selection practices while mitigating the impact of human biases such as in-group favoritism~\cite{masuda2015evolutionary,chang2021elitism}. 

In sum, we calculate the probability of hiring any of the top $d$ candidates using multiple stopping criteria and with closed form solutions. To avoid extensive computational time, we use the simulated annealing method for optimization, which performs reasonably well and saves significant computing time compared to exhaustive search. Our findings highlight the importance of using adaptive search strategies and computational methods for maximizing the likelihood of suboptimal outcomes in the satisficing secretary problem.

\section*{Acknowledgments}
F.F. gratefully acknowledges support from the Bill \& Melinda Gates Foundation (award no. OPP1217336), the NIH COBRE Program (grant no.1P20GM130454), and the Neukom CompX Faculty Grant. Roberto Brera acknowledges Ian Gill's significant contributions in the project's initial phase.

\end{document}